\documentclass[preprint]{aastex}
\usepackage[utf8]{inputenc}
\usepackage[margin=1.0in]{geometry}
\usepackage{amsmath} 
\usepackage{verbatim}
\usepackage{graphicx}
\usepackage{graphics}
\usepackage{url}

\title{UrHip Proper Motion Catalog}
\author{J. Frouard, B.N. Dorland, V.V. Makarov, N. Zacharias, C.T. Finch}
\affil{United States Naval Observatory, Washington, DC 20392, USA}
\email{jfrouard@federatedit.com}

\begin{abstract}
Proper motions are computed and collected in a catalog using the Hipparcos positions (epoch 1991.25) and URAT1 positions (epoch 2012.3 to 2014.6). The goal is to obtain a significant improvement on the proper motion accuracy of single stars in the northern hemisphere, and to identify new astrometric binaries perturbed by orbital motion. For binaries and multiple systems, the longer baseline of Tycho2 ($\sim$ 100 yr) makes it more reliable despite its larger formal uncertainties. The resulting proper motions obtained for 67,340 stars have a consequent gain in accuracy by a factor of $\sim$ 3 compared to Hipparcos. Comparison between UrHip and Hipparcos shows that they are reasonably close, but also reveals stars with large discrepant proper motions, a fraction of which are potential binary candidates.
\end{abstract}


\begin{document}

\maketitle

\section{Introduction}

The Hipparcos mission (ESA, 1997) provided accurate proper motions of bright stars with median formal errors at the 1 mas yr$^{-1}$ level \citep{petal1997}. The recently released URAT1 catalog\footnote{The URAT1 catalog is available at the CDS via \url{http://cdsarc.u-strasbg.fr/viz-bin/Cat?I/329}.} \citep{zaetal15,za15} contains the precise ICRF positions of 228,276,482 objects in the North Hemisphere at epochs ranging from 2012.3 to 2014.6. The number of objects per magnitude peaks at magnitude 18 in the 680-760 nm bandpass used (between R and I), and the median formal error in position for stars within the Hipparcos magnitude range (e.g. with completeness V = 7.3 - 9 and limiting V = 12.4) is 8 mas. URAT1 covers the -15$^o$ $\lesssim \delta <$ +90$^o$ sky area.

In this paper we combine the positions of Hipparcos and URAT1 to derive accurate proper motions, in much the same spirit as the proper motion catalog ACT \citep{uetal1998a}. The effective timespan from the two epochs is on average $\sim$ 22 years. 
This timespan allows us to improve on the formal uncertainty of the Hipparcos proper motions by approximately a factor
of 3 and to detect more accurately astrometric binaries, i.e., unresolved systems whose photocenter motion is
perturbed by orbital motion. A comparison between the new reduction of Hipparcos \citep{v2007a} and URAT1 positions was made by \cite{za15} in order to gauge the reliability of Hipparcos at the current epoch. About 20\% of the Hipparcos stars were found to have discrepant positions compared to URAT1 on a 3 sigma level (75 mas or more), likely due to inaccurate proper motions and binary motions in Hipparcos.

The objective of this proper motion catalog is two-fold. First, we aim at obtaining very accurate proper motions at a fraction of a mas yr$^{-1}$ level and thus gain a significant increase in accuracy compared to Hipparcos. The mean epoch of
Hipparcos measurements is J1991.25. We find in a number of applications, where the apparent places of reference stars
are needed for the current epoch, that the current positions of stable, single stars has degraded to $\sim 25$ mas
per coordinate or more due to the accumulated proper motion error. This puts Hipparcos on the same tier with formally
less precise, but much larger, ground-based catalogs such as UCAC4 \citep{za13} or URAT1. On the other hand,
Hipparcos remains the basic realization of the Celestial Reference Frame (CRF) in the optical band, and any fractional
improvement of Hipparcos proper motions results in a systematic improvement of the up-to-date CRF.

Secondly, this catalog will provide a check and constraints on unknown astrometric binaries. The Tycho-2 proper motions \citep{hetal2000,uetal2000} can be considered more reliable than the majority of astrometric catalogs when it comes to binary motions. Indeed the $\sim$ 100-year timespan used in the computation of the Tycho-2 proper motions, implied by the use of transit circles, photographic catalogs, the Astrographic Catalog \citep{uetal1998b}, and the Tycho-2 positions, has the effect of averaging out the orbital motions of binaries. Statistically significant differences between the short-term
Hipparcos and long-term Tycho-2 proper motions, therefore, indicate astrometric binaries with detectable
orbital motion. Using this method, \citet{mk2005} detected $\sim$ 2000 systems, while we detect an 
additional $\sim$ 5000 potential candidates from the difference between the UrHip and Hipparcos catalogs.

In Section 2, we present the methods we used to compute the UrHip proper motions. 
Section 3 is devoted to the results of this study, in particular in terms of accuracy, comparison with other proper motion catalogs, and distribution of binaries. Finally, the last section summarizes the features of our catalog and concludes this study. Efficient operations on astrometric catalogs were carried out on a PostgreSQL database with the Q3C sky indexing scheme \citep{kb06}.

\section{Methodology}

The Hipparcos catalog positions were propagated to the mean URAT1 epoch (2013.45) and cross-matched to the URAT1 positions with a 3 arcsec radius, providing 67,340 matched stars. When confronted with multiple matches within the search radius, we kept the closest one. Adding a criterion based on similar magnitudes for these multiple matches did not change the outcome. In principle, the Hipparcos proper motions may be perturbed by such a large amount ($>100$ mas yr$^{-1}$) that the existing URAT1 counterpart is missed, or an unrelated star is picked, but we consider this possibility to be unlikely because of the small number of nearby systems with large apparent motion. We also note that URAT1 only contains double stars if they have separations wider than several arcseconds, or if they appear as single image with the about 1 arcsec URAT1 resolution \citep{zaetal15}. The Hipparcos stars were then propagated from 2013.45 to the individual mean URAT1 epoch of their cross-matched URAT1 stars. 

With the goal of separating astrometrically stable stars from binaries, we first identified 10,151 stars flagged as known or suspected binaries in the Hipparcos catalog\footnote{Those have the h59 ``Multflg" flag with C,G,O,V,X values in the Hipparcos terminology.}. As a second step, we identified the Hipparcos astrometric binaries listed by \citet{mk2005} from our sample (specifically, the so-called $\Delta\mu$-binaries). Those are binaries detected from their discrepant proper motions between Hipparcos and Tycho-2. 1,483 of those entries already have a Hipparcos binary flag, while 647 do not. Concerning the precision of the cross-match, we note that, unsurprisingly, 259 out of the 265 stars outside a 1" radius are either astrometric binaries or have a h59 flag in Hipparcos (42 out of 43 stars for a 2" radius). Apart from Section 3.3 where we will investigate the distribution of those binaries, and unless stated otherwise, the rest of our analysis in this paper concerns the remaining 56,542 stars. 

Proper motions and their standard deviations were then computed from the sets of positions ($\alpha,\delta$) and their uncertainties ($\sigma_{\alpha},\sigma_{\delta}$) at epoch $t_H=1991.25$ (Hipparcos epoch) and $t_U =[2012.3-2014.6]$ (URAT1 epoch range) with the simple formulas
\begin{equation}
\mu_{\alpha^*} = \frac{\alpha_{_U} - \alpha_{_H}}{t_U - t_H} \cos \delta_{H} \text{ , \hspace{1cm}} \mu_{\delta} = \frac{\delta_{U} - \delta_{H}}{t_U - t_H} ,
\label{eq2}
\end{equation}
\begin{equation}
\sigma_{\mu_{\alpha^*}} = \frac{\sqrt{\sigma_{\alpha_{_U}}^2 + \sigma_{\alpha_{_H}}^2}}{t_U - t_H} \cos \delta_{H} 
= \frac{\sqrt{\sigma_{\alpha_{_U}^*}^2 + \sigma_{\alpha_{_H}^*}^2}}{t_U - t_H}  
\text{ , \hspace{1cm}} 
\sigma_{\mu_{\delta}} = \frac{\sqrt{\sigma_{\delta_{U}}^2 + \sigma_{\delta_{H}}^2}}{t_U - t_H} . 
\label{eq1}
\end{equation}
The effective timespan dictated by the sample has a minimum of 21.5 yrs, a maximum of 23.4 yrs and a median at 22.6 yrs.

\begin{figure}[h]
\resizebox{9cm}{!}{\includegraphics [angle=270,width=\textwidth] {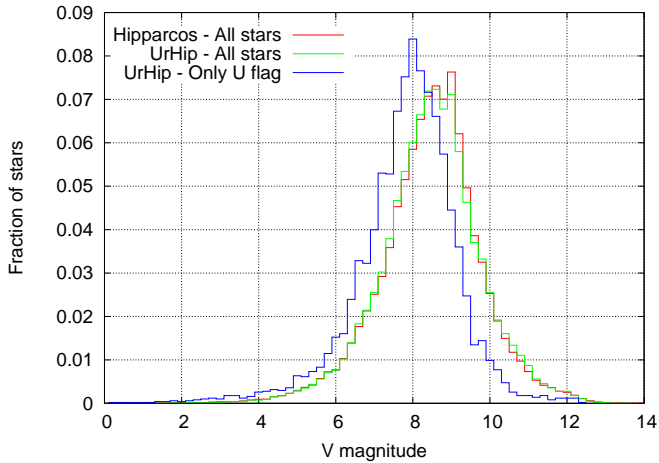}}
\caption{Fraction of stars in function of the Hipparcos V magnitude for the Hipparcos and UrHip catalogs, and for the subset of UrHip with the flag "U" (potential binaries - see section 3.3).}
\label{fig4}
\end{figure}

Table \ref{tab1} gives the median uncertainties in position and proper motion of the sample of 56,542 stars from the Hipparcos, URAT1 and UrHip catalogs, along with the uncertainties of the Tycho-2 stars having a Hipparcos counterpart, and the new reduction of Hipparcos \citep{v2007a}. The distribution of the Hipparcos V magnitude for Hipparcos and UrHip (including binaries) is shown in Fig.\ref{fig4}. The median V magnitude of the UrHip sample is 8.4 mag. The similar magnitude distribution for Hipparcos and UrHip shows the satisfactory completeness of URAT1 for bright stars. 
\begin{table}[h]
\begin{center}
\begin{tabular}{lllll}
	  & $\sigma_{\alpha^*} \text{(mas)}$ & $\sigma_{\delta} \text{(mas)}$ &$\sigma_{\mu_{\alpha^*}}$(mas yr$^{-1}$) & $\sigma_{\mu_{\delta}}$(mas yr$^{-1}$)\\
\hline
Hipparcos & 0.87 & 0.69 &1.02 & 0.82\\
URAT1 	  & 8 & 8 &5.6 &5.6 \\ 
\textbf{UrHip}	 & \textbf{-} & \textbf{-} &\textbf{0.35}&\textbf{0.35}\\
Tycho-2 &7 &8& 1.2&1.2\\
Hipparcos2 & 0.7 & 0.57 &0.85 & 0.7\\
\end{tabular}
\caption{Median uncertainties of the sample of 56,542 stars in different astrometric catalogs.\label{tab1}}
\end{center}
\end{table} 

\begin{figure}[h]
\resizebox{9cm}{!}{\includegraphics [angle=270,width=\textwidth] {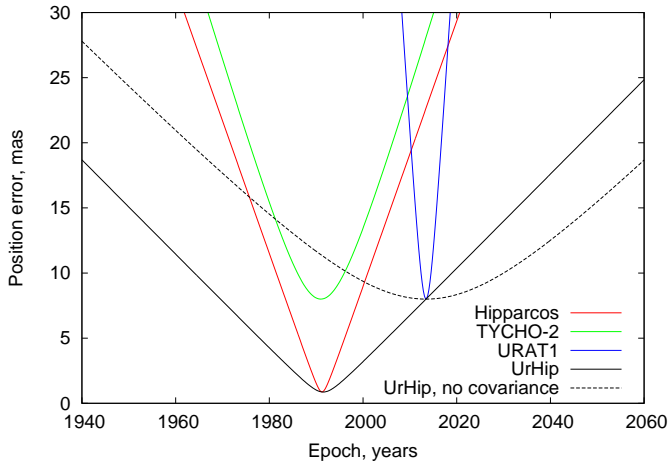}}
\caption{Propagation of the position errors (standard deviations) for UrHip, Hipparcos, Tycho-2 and URAT1.}
\label{fig5}
\end{figure}

Since we computed the proper motions from the positions (Eq.\ref{eq2}), we have to take into account the correlations of the proper motions with the positions in order to compute the propagation of the position errors. For a generic coordinate $p$, the position at epoch $t$ from the URAT1 position $p_{U}$ at epoch $t_{U}$ is 
\begin{equation}
p(t-t_{U}) = p_{U} + \mu_p (t-t_{U}). 
\label{eq3}
\end{equation}
The variance of the position is (see for example \cite{ks04})
\begin{equation}
\sigma^2_p(t-t_U) = \sigma_{p_U}^2 + \sigma_{\mu_p}^2 (t-t_U)^2 + 2 (t-t_U) \sigma_{p_U,\mu_p}
\label{eq4}
\end{equation}
where $\sigma_{p_U,\mu_p}$ is the covariance of $p_U$ and $\mu_p$, and can be computed as the expected value
\begin{equation}
\sigma_{p_U,\mu_p} = \mathrm{E}[(p_U - \bar{p_U})(\mu_p - \bar{\mu_p})] = \frac{\sigma_{p_U}^2}{t_U-t_H},
\end{equation}
where we used $\mu_p = \frac{p_U-p_H}{t_U-t_H}$, and we considered the URAT1 positions to be independent of the Hipparcos positions. The correct variance of the position errors are thus
\begin{equation}
\sigma_{\alpha^*}^2(t-t_U) =  \sigma_{\alpha^*_U}^2 + \sigma_{\mu_{\alpha^*}}^2 (t-t_U)^2 + 2 (t-t_U) \frac{\sigma_{\alpha_U^*}^2}{t_U-t_H} ,
\end{equation}
\begin{equation}
\sigma_{\delta}^2(t-t_U) =  \sigma_{\delta_U}^2 + \sigma_{\mu_{\delta}}^2 (t-t_U)^2 + 2 (t-t_U) \frac{\sigma_{\delta_U}^2}{t_U-t_H} .
\end{equation}
The minimum of the variance for a coordinate $p$ is 
\begin{equation}
\sigma^2_p(t^*-t_U) = \frac{\sigma^2_{p_H} \sigma^2_{p_U}  }{\sigma^2_{p_H} + \sigma^2_{p_U}}, 
\text{ reached at the epoch } 
t^* = \frac{t_H \sigma^2_{p_U} + t_U \sigma^2_{p_H}}{\sigma^2_{p_H} + \sigma^2_{p_U}}.
\end{equation}
Both are close to the Hipparcos values. The typical position errors for different catalogs over time are shown in Fig.\ref{fig5}. The propagation of the errors are computed by using the median values in Table \ref{tab1}. Taking into account the covariance term can lead to a $\sim$ 10 mas difference in estimating the UrHip error.

\section{Results}

The resulting UrHip catalog includes a number of data fields derived in this paper or copied from the source catalogs. The fields are: 
1) Hipparcos ID number; 
2) Tycho-2 ID number (concatenated from Tycho-2's TYC1, TYC2 and TYC3); 
3) right ascension in degrees from URAT1; 
4) declination in degrees from URAT1; 
5) formal error for URAT1 position; 
6) URAT1 epoch; 
7) UrHip proper motion in right ascension, in mas yr$^{-1}$ times $\cos\delta$; 
8) UrHip proper motion in declination, in mas yr$^{-1}$; 
9) standard error of proper motion in right ascension in field 7, in mas yr$^{-1}$ times $\cos\delta$; 
10) standard error of proper motion in declination in field 8, in mas yr$^{-1}$; 
11) binarity flag. 
The latter was copied from Hipparcos from the field h59 (C,G,O,V,X) whenever
present there, or set to ``A" for already identified astrometric binaries from \citep{mk2005}, or set to ``U" for
new suspected astrometric binaries as described in Section 3.3. A sample of the catalog is shown in the Appendix.

\begin{figure}[h]
\resizebox{9cm}{!}{\includegraphics [angle=270,width=\textwidth] {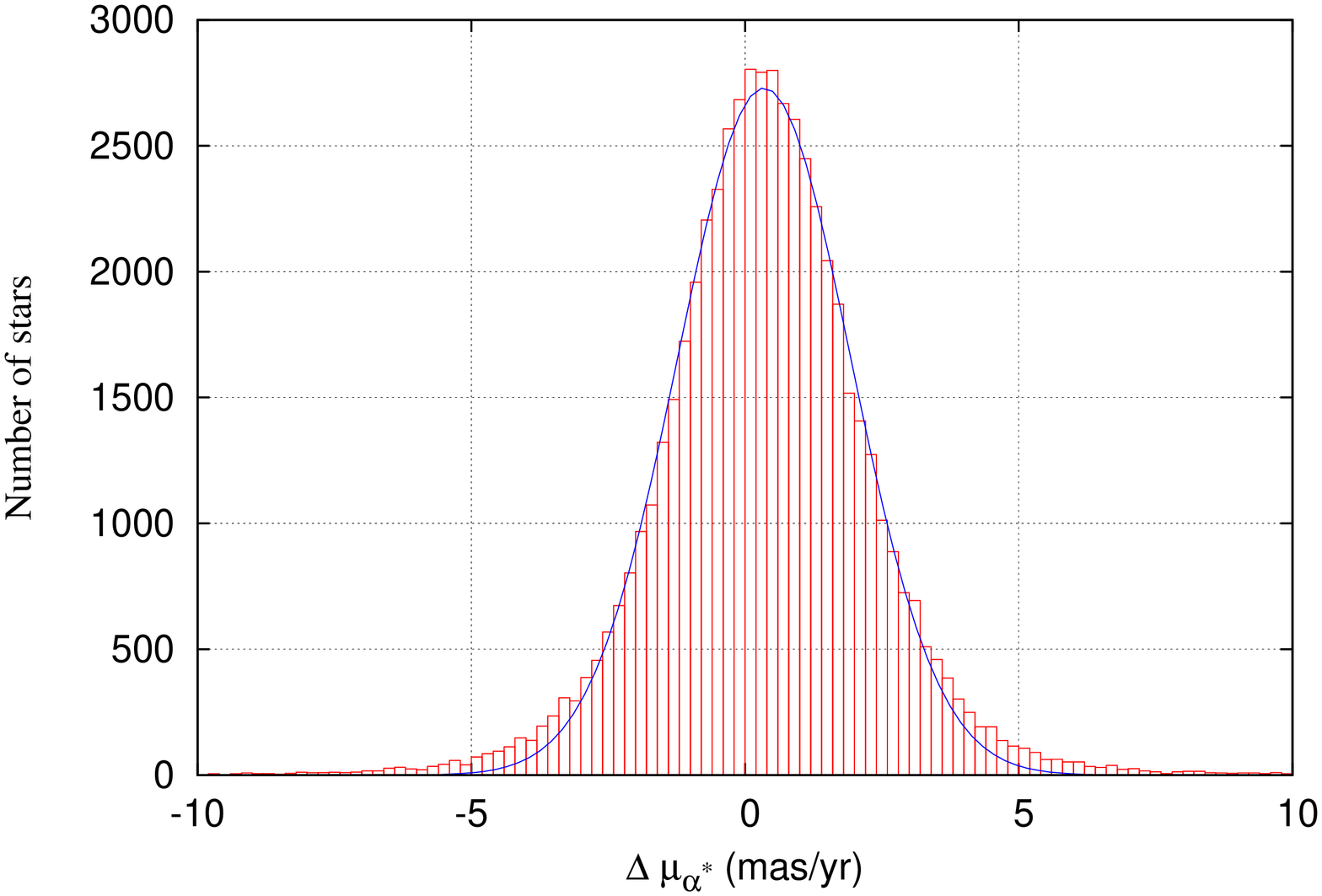}}
\resizebox{9cm}{!}{\includegraphics [angle=270,width=\textwidth] {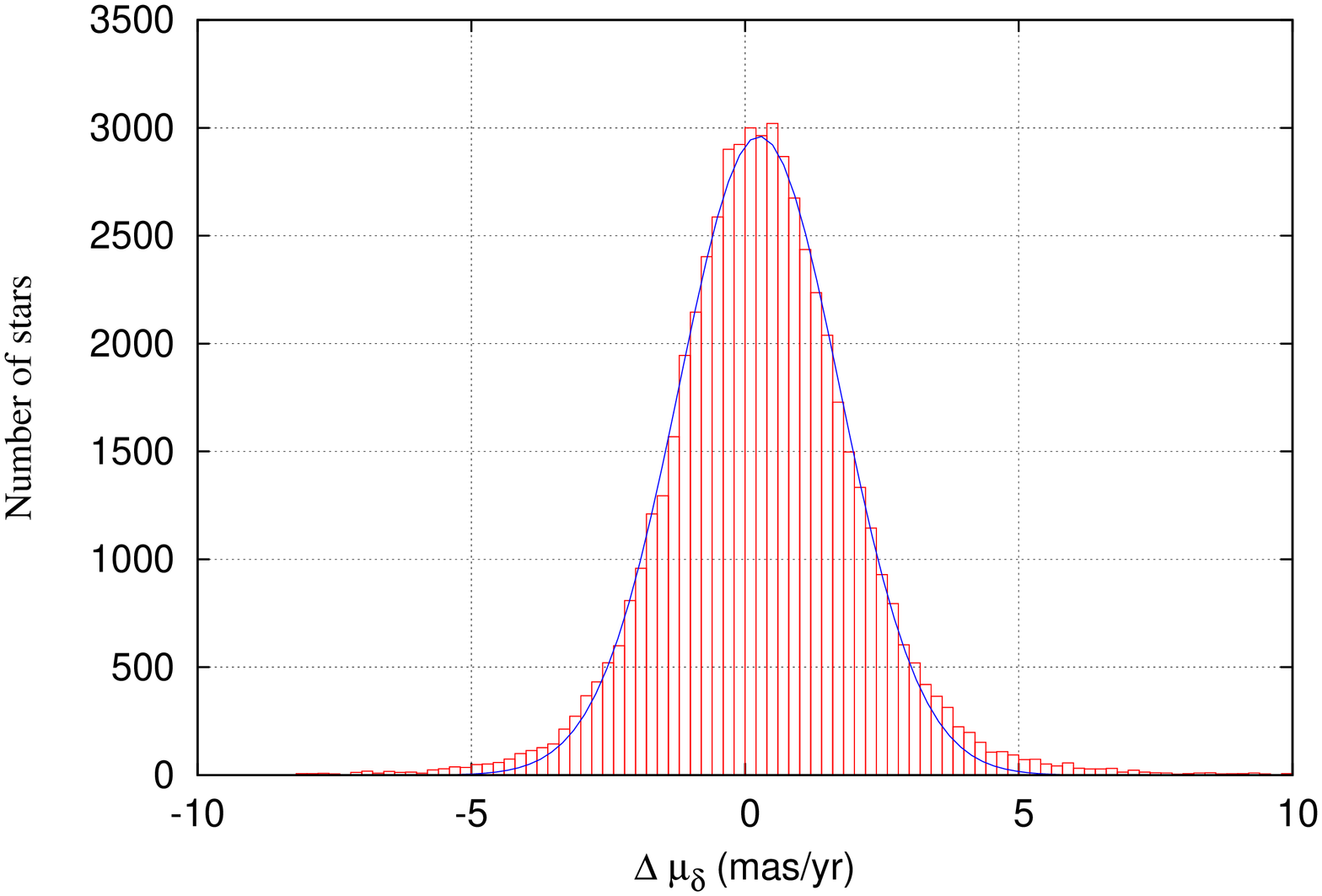}}
\caption{Differences in proper motion between UrHip and Hipparcos. The \{16\%, 50\%, 84\%\}  percentiles are \{-1.32, 0.351, 2.058\} mas yr$^{-1}$ in RA and \{-1.233, 0.276, 1.86\} mas yr$^{-1}$ in Dec. A gaussian fit gives a standard deviation = 1.584 mas yr$^{-1}$, mean = 0.350 mas yr$^{-1}$ in RA and a standard deviation = 1.471 mas yr$^{-1}$, mean = 0.261 mas yr$^{-1}$ in Dec.}
\label{fig1}
\end{figure}

\begin{figure}[h]
\resizebox{9cm}{!}{\includegraphics [angle=270,width=\textwidth] {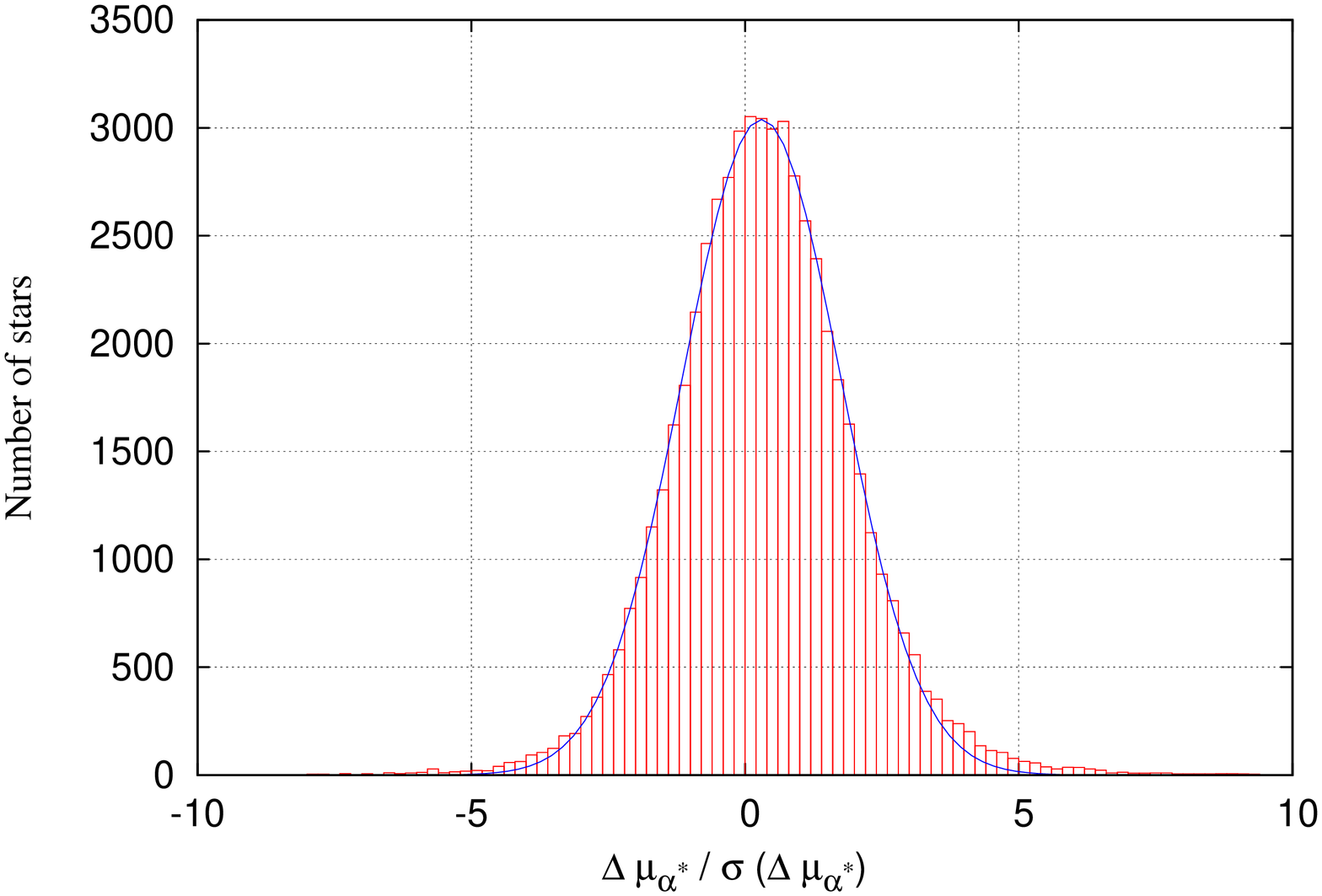}}
\resizebox{9cm}{!}{\includegraphics [angle=270,width=\textwidth] {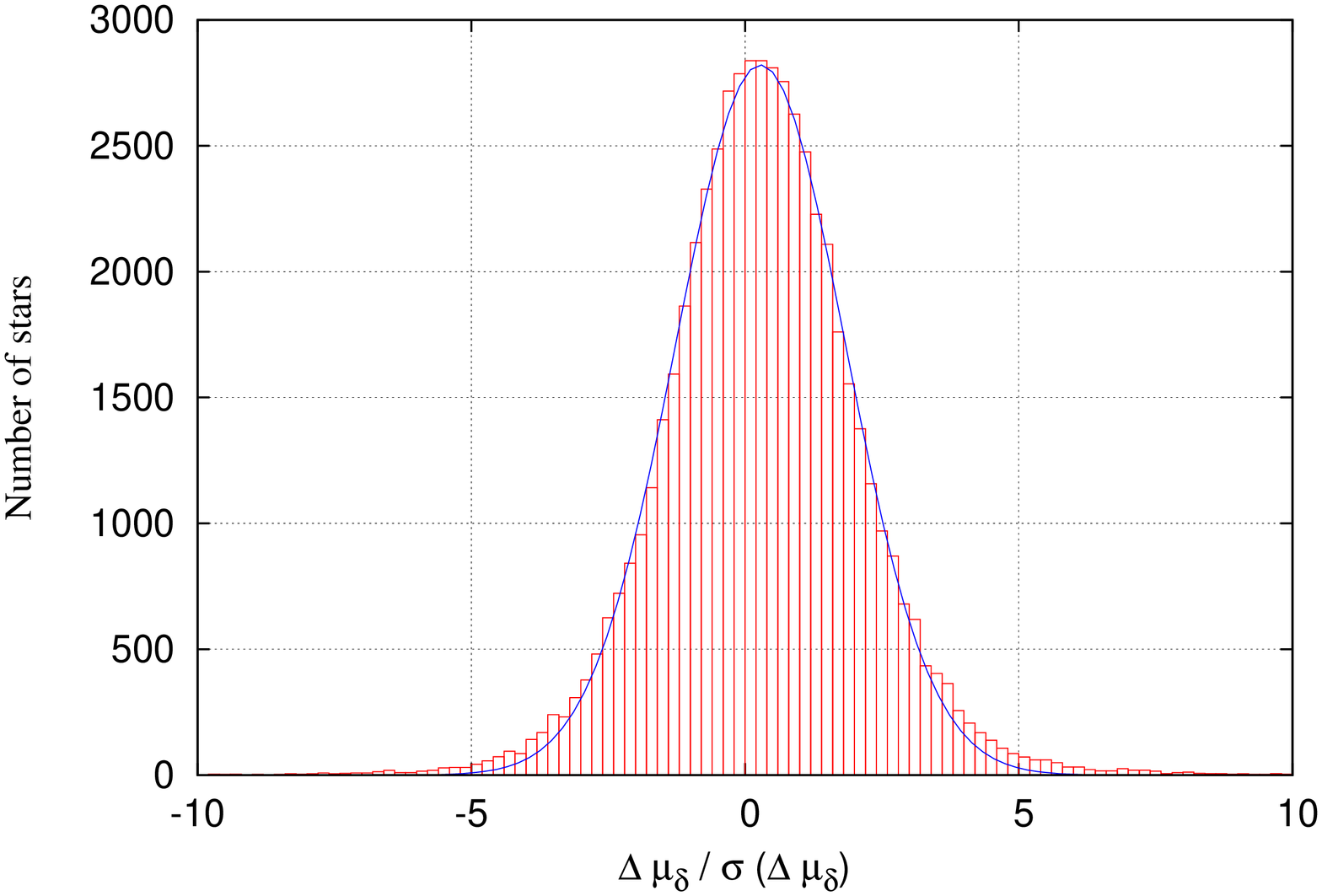}}
\caption{Differences in proper motion between UrHip and Hipparcos normalized by the combined standard error.  The \{16\%, 50\%, 84\%\}  percentiles are \{-1.151, 0.322, 1.854\} in RA and \{-1.312, 0.291, 1.92\} in Dec. A gaussian fit gives a standard deviation  = 1.448, mean = 0.303 in RA and a standard deviation = 1.553, mean = 0.284 in Dec.}
\label{fig2}
\end{figure}

The differences in proper motion (UrHip - Hipparcos) are shown in Fig.\ref{fig1}, while the differences normalized by the combined standard errors are shown in Fig.\ref{fig2}. A gaussian fit applied to the distributions (see Table \ref{tab2}) indicates that the formal errors have been underestimated, and is also a sign of an undetected fraction of binaries in the sample. From Fig.\ref{fig2} we note that $96\%$ of the stars in $\alpha^*$ and $94\%$ in $\delta$ are inside a 3.5 sigma limit. Some systematic offsets are present and correspond to 0.35 mas yr$^{-1}$ in $\mu_{\alpha^*}$ and 0.26 mas yr$^{-1}$ in $\mu_{\delta}$ (see Fig.\ref{fig1}). These small offsets are comparable to the deviation of the Hipparcos proper motions from the ICRS \citep{ketal1997}. They may also include small systematic errors in URAT1 positions. A number of UCAC4 stars \citep{za13} with good astrometry were used in the astrometric reduction of URAT1 observations. Although the intrinsic URAT1 system showed insignificant magnitude equation effects, magnitude-dependent offsets in UCAC4 could have affected URAT1 positions and introduced small sky-correlated errors. Most of the UCAC4 systematic errors are coma-like, i.e. depend on the product of magnitude and x,y coordinates on the CCD. Because the URAT1 field of view (2.65 by 2.65 deg per CCD) is much larger than the 1 deg FOV of UCAC, most of those errors will be averaged out in the URAT1 data. These errors are difficult to estimate, but are believed to be around 10 mas \citep{zaetal15}. Taking into account this systematic error to compute the uncertainties in Eq.\ref{eq1} would still grant a better proper motion accuracy than Hipparcos.

No significant systematic effects are observed when the proper motion differences are plotted against the Hipparcos V magnitude. The standard deviation of the proper motion differences is in close agreement to the Hipparcos - Tycho-2 value of 1.6 mas yr$^{-1}$ obtained by \cite{uetal2000}. 
 
\begin{table}
\begin{scriptsize}
\begin{center}
\begin{tabular}{l|ll|ll}
				&\multicolumn{2}{c|}{$\alpha^*$}		&\multicolumn{2}{c}{$\delta$}\\			
\hline
				& $\langle \Delta \mu \rangle$ & $\sigma (\Delta \mu)$ & $\langle \Delta \mu \rangle$ & $\sigma(\Delta)$\\
UrHip - Hipparcos (mas yr$^{-1}$)	& 0.350 & 1.584 & 0.261 & 1.471\\
UrHip - Tycho-2 (mas yr$^{-1}$) 	& 0.359 & 1.603 & 0.284 & 1.575\\
UrHip2 - Hipparcos2 (mas yr$^{-1}$) 	& 0.365 & 1.611 & 0.279 & 1.499\\
\hline
				& $\langle \Delta \mu / \sigma (\Delta \mu) \rangle$ & $\sigma (\Delta \mu / \sigma (\Delta \mu))$ & $\langle \Delta \mu / \sigma(\Delta \mu) \rangle$ & $\sigma(\Delta \mu / \sigma(\Delta \mu))$\\
UrHip - Hipparcos 		& 0.303 & 1.448 & 0.284 & 1.553\\
UrHip - Tycho-2 		& 0.274 & 1.286 & 0.221 & 1.237 \\
UrHip2 - Hipparcos2 		& 0.343 & 1.676 & 0.330 & 1.754
\end{tabular}
\caption{Mean and standard deviation ($\langle ... \rangle$ and $\sigma (...)$ respectively) obtained from a Gaussian fit of the $\Delta \mu$ and $\Delta \mu / \sigma (\Delta \mu)$ distributions.\label{tab2}}
\end{center}
\end{scriptsize}
\end{table} 

\begin{figure}[h!]
\resizebox{9cm}{!}{\includegraphics [angle=270,width=\textwidth] {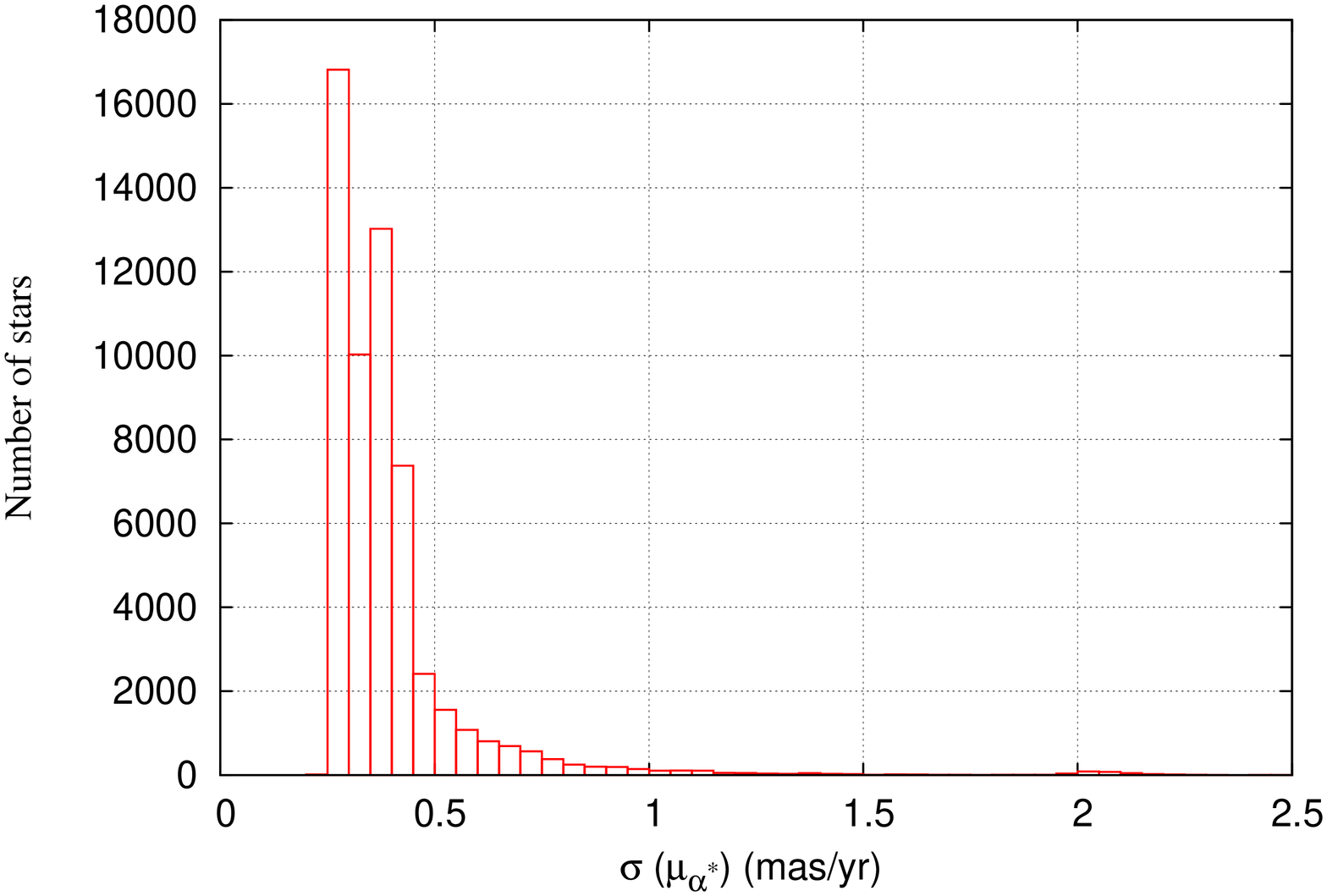}}
\resizebox{9cm}{!}{\includegraphics [angle=270,width=\textwidth] {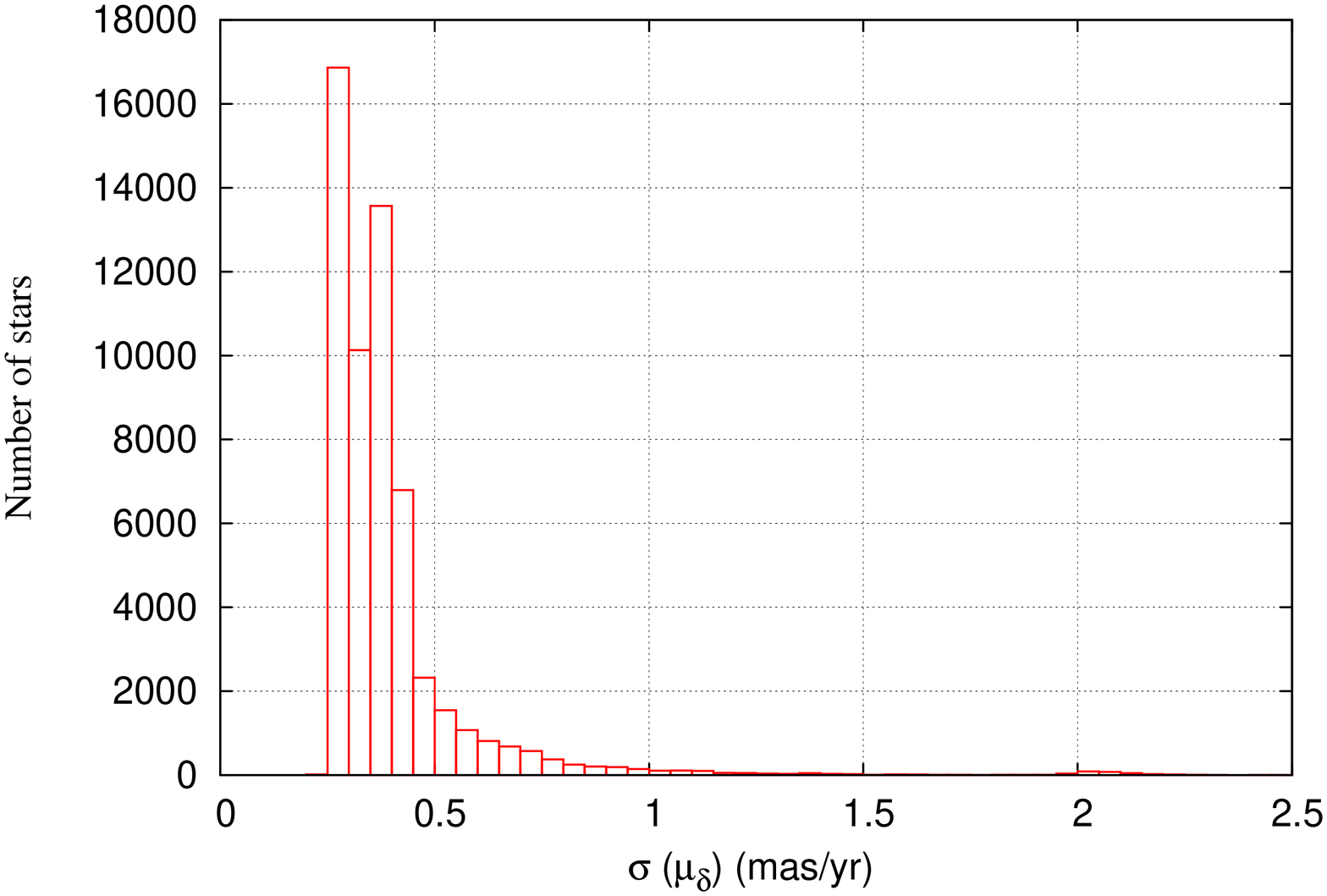}}
\caption{Distribution of UrHip $\sigma_{\mu_{\alpha^*}}$ and $\sigma_{\mu_{\delta}}$.}
\label{fig3}
\end{figure}

The distribution of UrHip proper motion uncertainties are shown on Fig.\ref{fig3} and their median values are given
in Table \ref{tab1}. Even though some large uncertainties are present, the bulk of errors is within 0.5 mas yr$^{-1}$,
reflecting a remarkable improvement with respect to the Hipparcos and Tycho-2 data (Table 1). A small secondary peak is present at $\sim 2$ mas yr$^{-1}$, which can be traced back to a small set of stars with larger URAT1 position errors of $\sim$ 45 mas, due to few observations available to compute the mean URAT1 positions. The median number of URAT1 observations used for the stars in our sample is 49, ensuring a median error of 8 mas.

Most URAT1 observations of Hipparcos stars were made with 10 and 30 sec exposures \citep{zaetal15}. For short exposures, the atmospheric turbulence induces some apparent displacements of the source (image motion effect, see \cite{l80,petal2002}) that short exposures cannot average out if the frequency spectrum of the perturbation contains long period terms. However, the large number of URAT1 observations per star helps in averaging out the effect, which is naturally taken into account in the URAT1 formal position errors. Another source of systematic errors is the atmospheric differential chromatic refraction (DCR) effect \citep{metal1992,setal2003,aetal2006} in which blue photons appear shifted toward the zenith compared to red photons. We expect this effect to be mitigated in URAT1 by the narrow bandpass of the filter used and the fact that all observations were restricted to within 20 minutes of the meridian.

\subsection{Comparison with Tycho-2 proper motions}

In this section we compare the UrHip and Tycho-2 proper motions. Their differences in proper motion are similar to Fig.\ref{fig1}-\ref{fig2} and their mean and standard deviations are shown in Table \ref{tab2}. 
The errors are similar (but slightly larger) to the differences UrHip-Hipparcos. In particular, the systematic deviations in both coordinates are in the same range. We advocate for the use of Tycho-2 proper motions
for flagged astrometric binaries (flags A and U). The longer baseline of Tycho-2 proper motions makes them more
stable in the presence of orbital motion as a higher degree of averaging is achieved.

\subsection{Using Hipparcos2 for the first epoch}

The same investigation can be made using the new reduction of the Hipparcos catalog \citep{v2007a,v2007b} (hereafter called Hipparcos2, while the new proper motions are called UrHip2) for the positions at the first epoch. The Hipparcos2 proper motions were used to propagate the Hipparcos2 stars to the mean URAT1 epoch. The cross-match is very similar to the Hipparcos-URAT1 cross-match described in Section 2 : 67,336 stars make the 3" radius cut, 12 stars are unique to the Hipparcos-URAT1 cross-match, while 8 stars are unique to the Hipparcos2-URAT1 cross-match. These 20 stars are all flagged as binaries in Hipparcos. The formal errors of the Hipparcos2 catalog for our sample, once the known and suspected binaries have been removed, are shown in Table \ref{tab1}. The differences in proper motion between UrHip2 and Hipparcos2 are shown in Table \ref{tab2}. The UrHip proper motion uncertainties are very similar when we chose Hipparcos2 instead of the original Hipparcos for the first epoch, because the uncertainties are mainly dependent on the URAT1 formal errors in position, which are larger than the Hipparcos and Hipparcos2 uncertainties by an order of magnitude (see Table \ref{tab1}). We conclude from this external check on proper motions, that the astrometric quality of Hipparcos2 and original Hipparcos
are practically the same.

\subsection{Large proper motion differences}

In the previous sections we did not investigate the distribution of known or already suspected binaries. As expected, those have significantly larger proper motion differences between UrHip and Hipparcos. $9\%$ of stars not yet recognized as binaries are outside of $\Delta \mu / \sigma (\Delta \mu) = 3.5$ for at least one coordinate, while this is true for $33 \%$ of the binaries flagged in the Hipparcos catalog, and $79 \%$ of the astrometric binaries found by \cite{mk2005} and not flagged in Hipparcos. Another way of stating these differences is to recognize that among all the stars located outside the $3.5 \sigma$ limit, $56 \%$ have not been detected as binaries in the past, while $38 \%$ are flagged in the Hipparcos catalog, and $6 \%$ are astrometric binaries found by \cite{mk2005} and not flagged in Hipparcos.

Now, discarding those known or suspected binaries, we still expect that many stars having large discrepant proper motions between UrHip and Hipparcos are likely to be unknown binaries. We count 221 stars with discrepant proper motion larger than 10 mas yr$^{-1}$ in at least one coordinate, while this number rises to 1997 for a 5 mas yr$^{-1}$ limit. We decided to give the flag ``U" in the catalog for all the stars having a proper motion difference normalized by the combined errors larger than 3.5 sigma in at least one coordinate, and not a previously known or suspected binary. This conservative limit yields 5054 flagged stars. Out of those 5054 flagged stars, we find 4213 which are also discrepant in a similar comparison of UrHip2-Hip2 and 1096 in a UrHip-Tycho2 comparison. The Hipparcos V magnitude distribution of those 5054 stars is shown on Fig.\ref{fig4}. The fact that those potential binaries can be detected implies that they are less distant (explaining the magnitude difference on Fig.\ref{fig4}), a fact which is confirmed by their larger parallaxes.
To summarize the binarity flag in the catalog:
\begin{itemize}
 \item 51488 stars (76 $\%$) do not have any flags,
 \item 2130 stars (3 $\%$) are recognized astrometric binaries: flag A, which supersedes any additional Hipparcos C,G,O,V,X flag,
 \item 8668 stars (13 $\%$) have a binarity flag in the Hipparcos catalog: flag C,G,O,V,X,
 \item 5054 stars (8 $\%$) are not known or suspected binary stars, but have discrepant proper motions between UrHip and Hipparcos: flag U.
\end{itemize}

\section{Conclusions}

In this paper we take advantage of the precise URAT1 and Hipparcos positions to compute accurate proper motions with a $\sim$ 22 years baseline. The gain in accuracy is significant: the UrHip formal errors are on average three times smaller than the Hipparcos formal errors. The catalog lists known detected and suspected binaries, but is still likely to contain unknown astrometric binaries which will be revealed by comparing UrHip proper motions with short-term Gaia \citep{p2003} proper motions. Taking advantage of the higher precision of UrHip proper motions, we detected $\sim$ 5000 stars, of which many are likely to be astrometric binaries.

\appendix
\section{Appendix. Catalog sample.}

\begin{scriptsize}

\begin{center}
\begin{tabular}{rlrrrlrrlll}
Hip & Tycho-2 & RA  & Dec  & $\sigma$(RA, Dec) & Epoch & $\mu_{\alpha^*}$ &  $\mu_{\delta}$  & $\sigma(\mu_{\alpha^*})$ & $\sigma(\mu_{\delta})$ & flag \\
 &  & deg & deg & mas & year & mas yr$^{-1}$ & mas yr$^{-1}$ & mas yr$^{-1}$ &mas yr$^{-1}$& \\
\hline
1&0001003811  &0.00088972&1.08898222&10&2013.617&-3.561&-5.006&0.451&0.448& \\
3&2781018631  &0.00506611&38.85926639&9&2013.554&7.310&-3.178&0.404&0.404&  \\
6&&0.01949639&3.94640889&9&2013.073&222.987&-13.204&0.452&0.424& \\
7&1184012071  &0.02115667&20.03537806&9&2013.586&-210.812&-197.294&0.405&0.404&U \\
8&1732027301  &0.02741639&25.88642250&12&2013.822&17.906&-8.285&0.537&0.533& \\
9&2271014111  &0.03531028&36.58599667&8&2013.569&-4.094&9.500&0.361&0.359& \\
11&3250008581  &0.03741306&46.93999222&9&2013.547&12.800&-1.505&0.404&0.404& \\
14&4663001601  &0.04864750&-0.36049333&14&2013.957&59.549&-11.437&0.618&0.617& \\
15&3258006601  &0.05043861&50.79120500&7&2013.653&13.176&5.007&0.314&0.314& \\
... &  &  &  &  &  &  &  &  && \\
120005&3806018191  &138.59341194&52.68398528&16&2013.083&-1570.993&-657.227&1.184&0.978&C \\
120082&3452018351  &174.95791111&45.15658056&6&2013.808&16.406&-5.313&0.268&0.267&U \\
120148&1625009111  &300.75460722&20.09708639&9&2013.702&181.829&-11.926&0.407&0.407& \\
120155&2684003841  &305.38220306&36.92020694&7&2013.691&-1.135&-3.422&0.314&0.315& \\
120248&4109004771  &102.46454778&66.35783472&6&2013.626&200.802&-88.916&0.275&0.282& \\
120250&5193005801  &317.50790583&-1.86463500&11&2013.618&229.562&-27.341&0.502&0.501& \\
120276&&150.52326806&79.70550278&6&2013.890&156.059&-123.982&0.358&0.372& \\
120290&2598013511  &254.42081806&35.26971500&6&2013.674&18.493&0.440&0.284&0.286& \\
120313&1463003931  &206.39867806&17.74242500&6&2013.989&-2.233&-2.729&0.321&0.326&
\end{tabular}
\end{center}
\end{scriptsize}

\acknowledgments
The authors would like to thank the anonymous reviewer for helpful comments that helped improve the quality of this paper.

\end{document}